\documentclass[
amsmath,amssymb,reprint,superscriptaddress,prapplied,aps,
]{revtex4-2}

\usepackage{graphicx}
\usepackage{dcolumn}
\usepackage{bm}
\usepackage[colorlinks=true, allcolors=blue]{hyperref}

\usepackage[mathlines]{lineno}
\usepackage{xcolor}
\usepackage{orcidlink}

\usepackage{lipsum}
\usepackage{float}

\begin{document}

\preprint{APS/123-QED}

\title{Harnessing bound states in the continuum for quantum and nonlinear photonics in silicon nitride microresonators}

\author{Eduardo C. Lima\orcidlink{0000-0003-0239-0340}}
 \email{lima.eduardo@ufabc.edu.br}
 \affiliation{Centro de Ci\^{e}ncias Naturais e Humanas, Universidade Federal do ABC-UFABC, Santo Andr\'{e} 09210-580, Brazil}
 \affiliation{Dipartimento di Ingegneria Industriale e dell’Informazione, Università di Pavia, Via Ferrata 5, 27100 Pavia, Italy.}
 
\author{Marco Clementi\orcidlink{0000-0003-4034-4337}}%
\affiliation{%
 Dipartimento di Fisica “A. Volta", Università di Pavia, Via Bassi 6, 27100 Pavia, Italy.
}%
\author{Alice Viola\orcidlink{0009-0001-6740-4195}}
\affiliation{%
 Dipartimento di Fisica “A. Volta", Università di Pavia, Via Bassi 6, 27100 Pavia, Italy.
}%
\author{Matteo Galli\orcidlink{0000-0001-7481-4806}}
\affiliation{%
 Dipartimento di Fisica “A. Volta", Università di Pavia, Via Bassi 6, 27100 Pavia, Italy.
}%
\author{Daniele Bajoni\orcidlink{0000-0001-6506-8485}}
 \affiliation{Dipartimento di Ingegneria Industriale e dell’Informazione, Università di Pavia, Via Ferrata 5, 27100 Pavia, Italy.}

 \author{Marco Liscidini\orcidlink{0000-0003-4001-9569}}
\affiliation{%
 Dipartimento di Fisica “A. Volta", Università di Pavia, Via Bassi 6, 27100 Pavia, Italy.
}%
 
\author{Massimo Borghi\orcidlink{0000-0003-4137-0852}}
\email{massimo.borghi@unipv.it}
\affiliation{%
 Dipartimento di Fisica “A. Volta", Università di Pavia, Via Bassi 6, 27100 Pavia, Italy.
}%

\date{\today}

\begin{abstract}

\noindent
In this work we theoretically and experimentally investigate bound states in the continuum (BICs) in a reconfigurable integrated photonic structure, focusing on spontaneous four-wave mixing. We consider a configuration in which only one mode, the idler, is tuned to the BIC condition, and show that suppressing the radiative loss of this single mode profoundly reshapes the overall nonlinear dynamics. In particularly, we realize electrically tunable Friedrich-Wintgen BICs with \(Q>10^6\) in an interferometrically coupled silicon-nitride microresonator. Although inaccessible through linear excitation, the BIC is populated by four-wave mixing and read out through its radiative photon partner.  In the low-gain regime, we investigate the spectral, correlation, and coherence properties of the generated photon pairs, revealing distinctive signatures in the signal-idler correlations and an extracted signal whose coherence time approaches the intrinsic-loss-limited lifetime of the dark idler. In the high-gain regime, making one mode participating in optical parametric amplification a BIC enhances the parametric gain by 12 dB and lowers the oscillation threshold by 1.6 dB. These results establish BICs as a tool for mode-selective lifetime engineering, enabling control over both biphoton wavepackets and the threshold dynamics of integrated parametric devices.

\end{abstract}

\maketitle

\section{Introduction}

\noindent
First proposed in 1929 as a discrete and stationary energy eigenvalue surrounded by a continuum of solutions in a quantum potential well \cite{von1929some}, bound states in the continuum (BIC) are exceptional states whose stationary property corresponds to a mode with ideally no dissipation. 
One important class comprises symmetry-protected BICs \cite{khaliq2026recent}, which arise when the spatial symmetry of the bound mode is orthogonal with that of the available radiation channels, and consequently, radiative coupling is forbidden by the zero-resulting overlap integral. 

Friedrich-Wintgen BICs  are another relevant class of BICs that do not rely on symmetry mismatch but rather originate from the destructive interference of multiple decay paths from a bound mode coupled to a common radiation channel \cite{friedrich1985interfering}. 
In contrast with symmetry-protected BICs, which live in infinitely extended periodic structures, Friedrich-Wintgen BICs can be supported by  systems of finite size, making them appealing for practical implementations. 
Similarly, Fabry-Pérot BICs are based on two identical resonances from two interacting cavities coupled through a single radiation channel \cite{ndangali2010electromagnetic}.
BICs have been successfully translated into numerous systems which support wave dynamics. 
In the optical domain, BICs have been engineered across a wide array of architectures, including bulk photonic devices \cite{hwang2021ultralow}, metasurfaces \cite{ha2018directional,zong2023merging, ren2025dynamically} and guided wave integrated photonic devices \cite{yu2019photonic,zhou2025efficient,ye2024integrated}.
The very high (theoretically infinite) Q-factor can drastically modify light-matter interactions, making them attractive for applications in both classical and quantum optics. 
In these contexts, BICs have been successfully proven to be useful for a wide variety of applications, which include among the others low-power lasing \cite{yang2021low}, hyper-parametric oscillation \cite{lei2023hyperparametric}, sum-frequency generation \cite{ye2025sum}, second-harmonic generation \cite{wang2020doubly}, four-wave mixing \cite{liu2023high,moretti2024si},  on-chip color routing  \cite{shi2026chip}, label-free biosensing \cite{yesilkoy2019ultrasensitive}, temperature monitoring \cite{xu2020silicon} and wave-front control \cite{kang2022coherent}. 
We direct the reader to Refs. \cite{azzam2021photonic, khaliq2026recent, xu2023recent} for comprehensive reviews covering fundamental concepts and applications.\\
Recently, the use of BICs has been extended to the realm of quantum nonlinear optics, with theoretical proposals for enhancing photon-pair generation by spontaneous parametric down conversion (SPDC) \cite{zhong2025enhanced} in thin film lithium niobate dielectric metasurfaces, improving the single-photon purity and brightness through Kerr nonlinearites \cite{deng2025enhancement}, and experimentally demonstrating a BIC-assisted photon-pair source based on SPDC in a thin-film lithium niobate waveguide \cite{ye2024integrated}.
However, the generation of nonclassical light, such as energy-time correlated photon-pairs \cite{wen2023polarization} and squeezed states \cite{jia2026monolithic}, through BIC-assisted third-order ($\chi^{(3)}$) nonlinearities in low-loss integrated platforms such as silicon nitride (SiN) remains largely unexplored. 
To date, experiments involving SiN resonators and BICs have focused primarily on above-threshold parametric oscillation. 
In the classical regime, BICs have been shown to provide additional control over conversion efficiency by suppressing mode competition \cite{lei2023hyperparametric} and to enhance pump conversion efficiency in Kerr-soliton microcombs \cite{ChavezBoggio2022}.\\
In this work, we present an integrated SiN photonic circuit exhibiting Friedrich-Wintgen bound states in the continuum with quality factors exceeding $10^6$, and exploit them for BIC-assisted photon-pair generation via spontaneous four-wave mixing (SFWM). 
The device consists of a high-$Q$ SiN microresonator (the \textit{main resonator}) coupled at two points to an U-shaped access waveguide, thereby forming an asymmetric Mach-Zehnder interferometer (MZI) \cite{Paula2025,borghi2024}. 
By controlling the phase accumulated in the MZI, the effective coupling coefficient can be continuously tuned. 
For specific phase values, the net radiative coupling to the bus waveguide vanishes because the different leakage channels interfere destructively, giving rise, in the limit of perfect decoupling, to a Friedrich-Wintgen BIC. 
Selective control over the resonances at which this condition occurs is achieved by embedding an overcoupled \textit{auxiliary} (Aux.) resonator in one arm of the MZI, which acts as a resonant phase shift element, introducing a strong spectral dependence of the effective coupling coefficient in the vicinity of its resonances.

By using this highly reconfigurable device, we investigate the role of BICs in SFWM, examining how tuning one of the interacting cavity modes into the BIC condition influences the properties of the generated photon pairs and how the characteristics of the photon emitted into the dark BIC mode can be inferred from measurements of its twin photon radiatively coupled to the bus waveguide.
Furthermore, we explore the transition from quantum to classical nonlinear optics by investigating optical parametric amplification (OPA) of a weak signal seed when the idler mode is engineered as a BIC. 

The remainder of the paper is organized as follows. Section~\ref{sec:theory} presents the quantum optical description of BICs in our photonic device and highlights their distinctive properties. 
In Section~\ref{subsec:probe}, we describe the device architecture and operating principle and present experimental evidence for the formation of BICs. 
Section~\ref{subsec:BIC-SFWM} reports the experimental investigation of SFWM, including measurements of the single-photon count rates, coincidence rates, and the second-order auto- and cross-correlation functions of the signal and idler fields as one of the resonances is tuned into a BIC. 
Finally, Section~\ref{subsec:OPA} presents the results on optical parametric amplification, comparing the performance of the conventional and BIC-assisted configurations.
All the experimental observations are supported by numerical simulations, which are presented in Appendices~\ref{apx:A} and \ref{apx:OPA}.

\section{BIC in a ring resonator, quantization and nonlinear Hamiltonian \label{sec:theory}}
\begin{figure}[t]
    \centering
    \includegraphics[width=\linewidth]{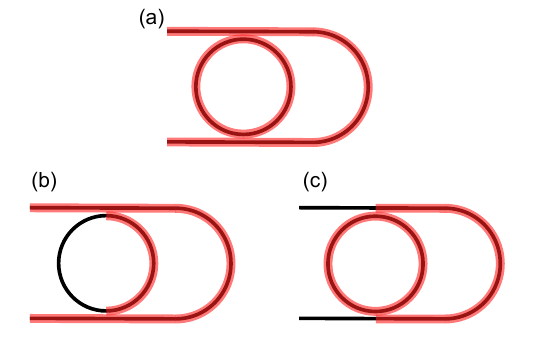}
    \caption{Schematic representation of the electromagnetic modes in a ring resonator coupled to a bus waveguide through an interferometric coupler. 
    (a) When the effective coupling between the bus waveguide and the ring is finite, Maxwell's equations admit only one solution at the resonant frequency $\omega_{\textrm{BIC}}$ involving both the bus waveguide and the resonator. 
    When the effective coupling vanishes at $\omega_{\textrm{BIC}}$, Maxwell's equations admit two distinct solutions at the same frequency. 
    One solution corresponds to a field propagating in the bus waveguide, which represents the continuum of modes (b). 
    The other is a bound solution whose field is confined within the ring resonator and does not radiate into the bus waveguide (c).}
    \label{fig:BIC_theo}
\end{figure}

\noindent
To illustrate the physical system presented here, we consider the structure shown in Fig.~\ref{fig:BIC_theo}, which can be regarded as a precursor to the one fabricated and studied in this work. 
The system consists of a ring resonator coupled to a bus waveguide through an interferometric coupler. 
This simplified structure allows us to better understand the physical mechanisms leading to BIC formation and to clarify the distinction between the case in which the system supports only a continuum of propagating modes and the case in which it also supports a BIC. 
This distinction is crucial for understanding the challenges associated with the quantization of the electromagnetic field and for the subsequent description of quantum nonlinear optical processes in our structure.\\
The MZI formed by the bus waveguide and the right side of the ring gives rise to an effective coupling between the ring resonator and the bus waveguide \cite{Paula2025,liu2019high}.
In the usual case shown in Fig.~\ref{fig:BIC_theo}(a), this coupling is finite at the resonant frequencies of the ring. 
The structure therefore behaves similarly to a conventional ring resonator coupled to a bus waveguide, with the main difference that the coupling is strongly frequency dependent. 
As a result, different resonant modes can exhibit different, but always finite, dwell times. 
At each resonant frequency, Maxwell's equations admit a single stationary solution of the type sketched in Fig.~\ref{fig:BIC_theo}(a), describing light entering and exiting the system through the bus waveguide and the field circulating in the resonator.
By contrast, the MZI can be designed so that, for a subset of resonant frequencies of the resonator, the effective coupling between the ring and the bus waveguide vanishes~\cite{ChavezBoggio2022}. 
When this occurs, Maxwell's equations admit two distinct solutions at the same frequency. 
One solution belongs to the continuum of asymptotic fields and describes light entering or leaving the structure through the bus waveguide, as sketched in Fig.~\ref{fig:BIC_theo}(b). 
The other solution is a square-integrable mode confined in the resonator, with no outgoing radiation toward the structure ports, as shown in Fig.~\ref{fig:BIC_theo}(c). 
This confined solution corresponds to a Friedrich-Wintgen  BIC ~\cite{friedrich1985interfering}. 
In this class of BICs, the state lies at a frequency within the radiation continuum of the access waveguide, but its net radiative amplitude vanishes because different leakage paths interfere destructively. 
At the BIC point, this cancellation suppresses radiation through all available external channels. 
Thus, in the ideal lossless limit, the radiative contribution to the linewidth is zero. 
In the presence of loss, the quality factor is instead limited by non-radiative or scattering loss mechanisms. 
Importantly, even when such losses are present, Maxwell's equations still admit two distinct solutions at the BIC frequency.\\
The coexistence of a propagating solution and a confined BIC mode has important consequences for quantization. 
Indeed, a complete expansion of the electromagnetic field cannot be constructed from the asymptotic states alone, because such a basis would not include the confined BIC mode. 
Therefore, one must start from the asymptotic-field formalism introduced for nonlinear electromagnetic Hamiltonians in Ref.~\cite{Liscidini2012} and extend the mode expansion by explicitly including the discrete BIC mode.
Within this framework, the displacement-field operator is thus written as
\begin{align}
\hat{\mathbf D}(\mathbf r) ={}&
\sqrt{\frac{\hbar\omega_{\mathrm{BIC}}}{2}}\,
\mathbf D^{\mathrm{BIC}}_{\omega_{\mathrm{BIC}}}(\mathbf r)
\hat b_{\omega_{\mathrm{BIC}}}
\nonumber\\
&+\int_{0}^{\infty} d\omega\,
\sqrt{\frac{\hbar\omega}{2v_g}}\,
\hat a_{\omega}\,
\mathbf D^{\mathrm{asy\text{-}in}}_{\omega}(\mathbf r)
+ \mathrm{H.c.},
\label{eq:displacement_bic}
\end{align}
where $\hat b_{\omega_{\mathrm{BIC}}}$ annihilates a photon in the BIC, $\hat a_\omega$ annihilates a photon in the asymptotic-in continuum, $v_g$ is the group velocity in the access waveguide, and $\mathbf D^{\mathrm{BIC}}_{\omega_{\mathrm{BIC}}}(\mathbf r)$ and $\mathbf D^{\mathrm{asy\text{-}in}}_\omega(\mathbf r)$ are the corresponding displacement profiles, solution of Maxwell's equations. 
The discrete BIC operator is dimensionless, whereas the continuum operator has units of $\mathrm{s}^{1/2}$, consistently with
\begin{equation}
[\hat b_{\omega_{\mathrm{BIC}}},\hat b_{\omega_{\mathrm{BIC}}}^{\dagger}]=1,
\qquad
[\hat a_{\omega},\hat a_{\omega'}^{\dagger}]=\delta(\omega-\omega').
\label{eq:commutation_continuum_bic}
\end{equation}
The BIC mode and the asymptotic fields are orthogonal. Therefore:
\begin{equation}
[\hat b_{\omega_{\mathrm{BIC}}},\hat a_{\omega}]
=[\hat b_{\omega_{\mathrm{BIC}}},\hat a_{\omega}^{\dagger}]=0,
\label{eq:bic_continuum_commute}
\end{equation}
even when $\omega=\omega_{\mathrm{BIC}}$. 
Such a description can also be generalized to include scattering losses \cite{banic2022two}.

In this work, we consider SFWM in the non-degenerate configuration, and we study the case in which one photon of the generated pair is emitted into a conventional radiative mode and the other is created in the BIC. 
Keeping the term that creates a continuum photon at frequency $\omega_1$ and a BIC photon at $\omega_2=\omega_{\mathrm{BIC}}$, while annihilating two pump photons from the continuum, the nonlinear Hamiltonian can be written as
\begin{align}
\label{eq:sfwm_bic_hamiltonian}
\hat H^{\mathrm{SFWM}}_{\mathrm{BIC}} ={}&
\int d\omega_1 d\omega_3d\omega_4\,
\\
&\times
S(\omega_1,\omega_{\mathrm{BIC}},\omega_3,\omega_4)
\hat a^{\dagger}_{\omega_1}
\hat b^{\dagger}_{\omega_{\mathrm{BIC}}}
\hat a_{\omega_3}
\hat a_{\omega_4}
+\mathrm{H.c.}\nonumber
\end{align}
\\
Here, $S(\omega_1,\omega_{\mathrm{BIC}},\omega_3,\omega_4)$ is a structure-dependent nonlinear coupling amplitude whose magnitude is determined by the spatial modal overlap of the modes involved in the nonlinear interaction mediated by the system’s third-order nonlinear response \cite{Liscidini2012}. 
Importantly, the integrations in Eq.~\eqref{eq:sfwm_bic_hamiltonian} extend only over the continuum of modes, while there is no integral over the BIC frequency. 
Thus, this term in the nonlinear Hamiltonian describes only SFWM in which one photon can be generated and remain trapped (in the absence of scattering losses) in the bound state, while its partner is emitted into the radiative mode of the bus waveguide.
As already mentioned, in the device studied experimentally below, the BIC is realized using a different but related architecture. 
The main ring is coupled to the bus waveguide through the MZI, in which an additional auxiliary ring resonator is overcoupled to the bus waveguide (see Fig.~\ref{fig:res}(a)). 
This auxiliary ring provides a narrow, resonance-dependent phase shift that allows one to control the effective coupling of a selected main-ring resonance ~\cite{Paula2025}. 
Naturally, from the perspective of each individual BIC, the theoretical description developed above remains valid. 
However, as we shall see below, the ability to isolate a single BIC has important consequences for controlling the nonlinear light-matter interaction.

\section{Experimental results \label{sec:exp}}
\begin{figure}[ht]
    \centering
    \includegraphics[width=\linewidth]{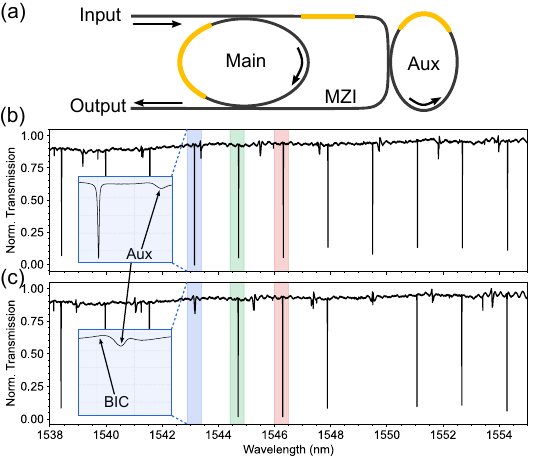}
    \caption{(a) Schematic of the photonic circuit under test. Black lines denote SiN waveguides, while the yellow regions correspond to metallic microheaters used for thermo-optic tuning of the device. Main: main resonator, Aux: auxiliary resonator, MZI: Mach-Zehnder interferometer. input and output denote the bus waveguides through which light is coupled into and out of the chip, respectively. (b) Normalized transmission spectrum of the device measured with no spectral overlap between the resonances of the main and Aux. resonators.
    (c) Normalized transmission spectrum for a configuration in which one out of every four resonances of the main resonator overlaps with a resonance of the Aux. resonator, creating BICs. The insets in (b) and (c) show magnified views of the transmission spectra in the vicinity of the main-resonator resonance highlighted in blue. The arrows indicate the resonance wavelength of the Aux. resonator and, in panel (c), the wavelength of the BIC mode. The three highlighted resonances are involved in the StFWM experiment described in the main text.
}
    \label{fig:res}
\end{figure}

\noindent
The device, sketched in Fig.~\ref{fig:res}(a), consists of a SiN main resonator with a free spectral range (FSR) of \mbox{$\thicksim195$ GHz}, coupled to the bus waveguide through a resonant interferometric coupler. 
This coupling scheme enables wavelength-selective tuning of the effective coupling coefficient between the main resonator and the input bus waveguide. 
The device design, geometry, and operating principle have been described in detail in Refs.~\cite{borghi2024, Paula2025}.

The FSRs of the MZI and the main resonator were designed to be nominally identical. 
In the absence of the Aux. ring, which is embedded in the longer arm of the MZI, this condition ensures that all resonances of the main resonator experience the same effective coupling coefficient. 
This coefficient can ideally be tuned from zero to $4\kappa^2(1-\kappa^2)$ (where $\kappa^2$ is the power coupling coefficient between the main ring and the bus waveguide, fixed by the coupling gap) by varying the phase accumulated in the longer arm of the interferometer using an integrated microheater \cite{Paula2025}.
The Aux. resonator introduces an additional phase shift for wavelengths close to its resonances, thereby modifying the relative phase between the two MZI arms and, consequently, the effective coupling coefficient between the main resonator and the bus waveguide. 
The FSR of the Aux. resonator is designed to be $\frac{4}{3}$ that of the main cavity. 
As a result, the effective coupling coefficient, and thus the extrinsic quality factor, of one out of every four resonances can be selectively tuned while leaving the remaining resonances essentially unaffected. 
The targeted resonances of the main resonator are selected by tuning the resonance wavelength of the Aux. cavity using an integrated thermo-optic phase shifter.
For this subset of resonances, fine adjustment of the resonance wavelength detuning, $\Delta\lambda_{\mathrm{aux}}$, between the main and Aux. resonators allows the effective coupling coefficient to be reduced to zero. 
Under this condition, the spatially localized mode supported by the main resonator and the interferometric coupler becomes decoupled from the bus waveguide, giving rise to a BIC whose lifetime is ultimately limited only by the intrinsic cavity losses.

\subsection{Creation and characterization of BIC states in the photonic device \label{subsec:probe}}
\noindent
To demonstrate the formation of BICs in our device, we first identified the electrical powers that must be applied to the integrated microheaters to achieve ideally zero net coupling between the main resonator and the access waveguide. Figure~\ref{fig:res}(b) shows the transmission spectrum of the device, measured by sweeping the wavelength of a continuous-wave (CW) laser injected into the input port and collecting the transmitted light at the output port (see Fig.~\ref{fig:res}(a) for ports description and position). This is accomplished by using an ultra-high numerical aperture (UHNA4) fiber array (coupling loss of $\thicksim3$ dB/facet) after that the polarization has been set to TE by means of a fiber polarization controller (FPC). The spectrum exhibits a double-periodic pattern of resonance dips. The shallower dips correspond to the resonances of the Aux. resonator, which is strongly overcoupled to the bus waveguide and exhibits a quality factor of approximately $4\times10^4$. The deeper dips correspond to the resonances of the main resonator. In this configuration, the loaded quality factor ($Q_{\textrm{L}}$) of the main resonator is set to approximately $1.3\times10^6$ by adjusting the electrical power applied to the microheater integrated in one arm of the MZI, and which corresponds to critical coupling. The intrinsic quality factor ($Q_{\textrm{int}}$) is then expected to be $Q_{\textrm{int}}=2Q_{\textrm{L}}=2.6\times10^6$.

\begin{figure*}[ht]
    \centering
    \includegraphics[width=\linewidth]{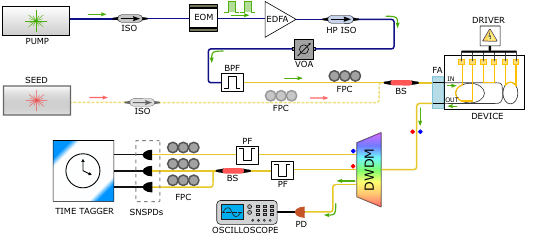}
    \caption{Sketch of the experimental setup used for the SFWM and StFWM characterization of the device under test. (HP) ISO: (high-power) fiber-optic isolator, EOM: electro-optic amplitude modulator, EDFA: Erbium-doped fiber amplifier, VOA: variable optical attenuator, BPF: bandpass filter, FPC: fiber polarization controller, BS: 50:50 fiber beam splitter, FA: fiber array, DWDM: dense wavelength division multiplexing module, PF: pump suppression filter, SNSPDs: superconducting nanowire single-photon detectors, PD: photodetector. The blue,  yellow and black paths refer to polarization maintaining fibers, single mode fibers, and electrical connections  respectively.}
    \label{fig:setup}
\end{figure*}

In the configuration shown in Fig.~\ref{fig:res}(b), the resonances of the main cavity do not overlap with those of the Aux. resonator. Consequently, all main-resonator resonances exhibit the same quality factor and extinction. By slightly varying the electrical power applied to the heater integrated on the Aux. cavity, its resonances are progressively brought into coincidence with one out of every four resonances of the main resonator. As the detuning $\Delta\lambda_{\mathrm{aux}}$ is reduced, the extinction of these selected resonances gradually vanishes (see Fig.~\ref{fig:res}(c) and the insets in Figs.~\ref{fig:res}(b,c), which show a magnified view of one of the targeted resonances). These resonant modes therefore become linearly decoupled from the continuum of modes of the bus waveguide; that is, they can no longer be excited by light injected into the input port at the corresponding BIC wavelength.\\

\begin{figure*}[!t]
    \centering
\includegraphics[width=1\linewidth]{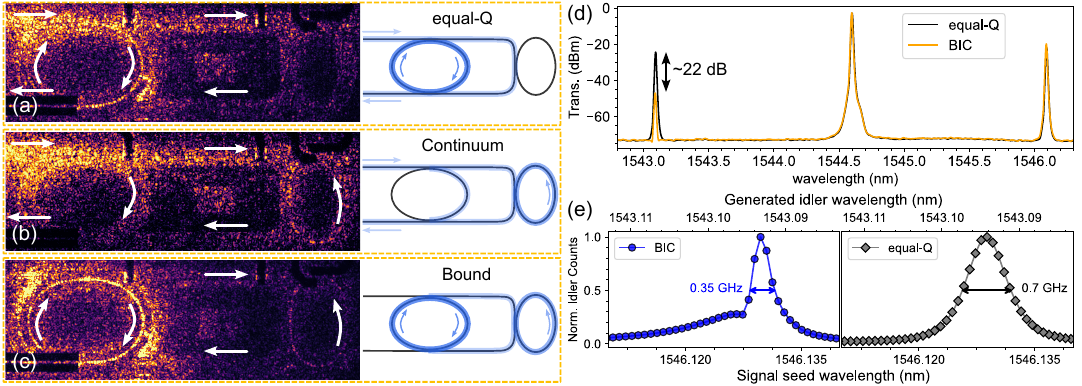}
    \caption{False color infrared camera images of the light scattered from the top of the photonic circuit and experimental results of StFWM experiments. (a) Idler-frequency light enters through the input waveguide and is resonantly coupled into the main resonator, which is configured to operate on critical-coupling conditions. (b) When the device supports a BIC at the input laser wavelength, the mode cannot be excited from the input bus waveguide, and consequently no resonant field enhancement is observed inside the main resonator. (c) Idler light generated inside the resonator through StFWM in the BIC configuration. This image has been acquired using a bandpass filter that transmits only the idler wavelength. The sketches on the right of each image illustrate the direction of light propagation in the device. (d) Spectrum of the transmitted light measured in the StFWM experiment for the \emph{equal-Q} and BIC configurations. The three peaks correspond to the pump (1544.6 nm), signal (1546.1 nm), and idler (1543.1 nm) beams involved in the StFWM process. (e) Intensity (scatter points) of the idler beam generated by StFWM using a $5$ ns pump pulse and a CW seed laser swept across the signal resonance. Light is collected at the output port of the device. The left and right panels correspond to the BIC and \emph{equal-Q} configurations, respectively. The solid lines are included only as guides to the eye.}
    \label{fig:stFWM}
\end{figure*}

To further validate this observation, we tuned the laser wavelength to one of the resonances supporting a BIC mode (highlighted in blue in the spectra shown in Figs.~\ref{fig:res}(b,c)) while simultaneously imaging the device from above using a $20\times$ microscope objective and an infrared camera. The scattered-light distribution obtained for the configuration shown in Fig.~\ref{fig:res}(b), in which the Aux. resonances are detuned from all main-ring resonances and therefore introduce no perturbation (hereafter referred to as the \emph{equal-Q} configuration), is shown in Fig.~\ref{fig:stFWM}(a) and reveals that light is circulating in the main resonator and in the bus waveguide. Although this is not immediately evident from the top-view image, the measured intensity distribution is consistent with the asymptotic input-field profile shown in Fig.~\ref{fig:BIC_theo}(a). Propagation of light through the bus waveguide from the input to the output port cannot be appreciated in the top-view image because of the weak radiative scattering from the waveguides and the absence of resonant field enhancement outside the microresonator. 
We then switched to the configuration shown in Fig.~\ref{fig:res}(c), supporting a BIC state at input laser wavelength, and injected light at the input port. The infrared image, shown in Fig.~\ref{fig:stFWM}(b), indicates that the light intensity inside the resonator has been  suppressed, consistently with the asymptotic field pattern sketched in Fig.~\ref{fig:BIC_theo}(b). As one can appreciate from the image, light is now also circulating inside the Aux. resonator, which has the active role of inducing the phase shift necessary for creating the BIC condition. This confirms that the BIC state cannot be externally excited from the input bus waveguide.\\ However, we now show that it could be internally excited through a nonlinear process, and we choose stimulated four-wave mixing (StFWM) as our probe mechanism. More in detail, we choose as the pump and the seed signal wavelength the resonances of the main resonator  highlighted in green and red respectively in Fig.~\ref{fig:res}(c). In this configuration, energy conservation implies the generation of a conjugate idler beam at the BIC wavelength. For this experiment, the pump intensity is amplified by an Erbium-doped fiber amplifier (EDFA) and a second CW seed laser is introduced and combined with the pump by a $50:50$ beam splitter (see Fig.~\ref{fig:setup}). We then placed a series of narrow-band bandpass filter, centered at the idler wavelength, along the microscope lens tube, in order to suppress the pump and signal light reaching the imaging sensor of the camera ($>80$ dB isolation).\\
When the seed laser is tuned into resonance, light is generated at the idler wavelength. The spectra of the transmitted intensity at the output port is recorded by an Optical Spectrum Analyzer (OSA).
The image in Fig.~\ref{fig:stFWM}(c), acquired under StFWM in the BIC configuration, shows the generated idler light circulating in the main resonator. A nearly identical top-scattering pattern (not shown) is observed when the device is operated in the \emph{equal-Q} configuration, indicating that the intracavity idler field remains essentially unchanged. The key difference lies in the extracted idler power: in the BIC configuration, the idler light coupled into the bus waveguide and collected at the output port is suppressed by more than two orders of magnitude ($\sim 22$ dB) compared with the \emph{equal-Q} configuration, as shown in Fig.~\ref{fig:stFWM}(d). 
This observation confirms that, in the BIC configuration, the escape efficiency of the idler photons into the continuum of modes of the bus waveguide is drastically reduced. It is worth noting that in an ideal BIC, this quantity must be zero. In practice, there is a very small residual coupling which is  caused by the finite extinction of the MZI coupler, most likely due to the non-identical point-couplers between the main resonator and the bus waveguide. Therefore, in the remainder of the paper, we will refer to the \emph{BIC condition} as the configuration at which the effective coupling with the bus waveguide is minimized. As shown later in Section~\ref{subsec:BIC-SFWM}, the residual coupling plays a negligible role in determining the lifetime of the BIC, which is instead dominated by intrinsic losses.
Spurious StFWM occuring along the input waveguide might contribute as well to the residual idler intensity in the output port.\\  
To further support the evidence that the BIC can be internally  excited through StFWM, we scan the seed laser across the signal resonance while keeping fixed the pump wavelength, and we monitored the intensity of the residual idler light leaked into the output port. 
To mitigate thermal drifts of the resonance wavelengths induced by the strong pump beam over the long duration of this measurement, where the seed laser is slowly tuned in step-mode using a piezo-electric actuator, the pump laser is modulated by an electro-optic amplitude modulator (EOM) at a repetition rate of $2$ MHz to create rectangular pulses with a duration of $5$ ns ($200$ MHz spectral width), which is sufficiently long to let the cavity to reach a steady-state, but short enough to avoid the build up thermal drifts, which require longer time scales \cite{brusaschi2026time}.
For this measurement, the Q-factors of the pump and signal resonances are set to $Q\thicksim 1\times10^5$ ($2$ GHz) for the BIC configuration and $Q\thicksim2\times10^5$ ($1$ GHz) for the \emph{equal-Q} configuration. In the \emph{equal-Q} configuration, the idler resonance at $\bar{\omega}_i$ share the same Q-factor of the pump and signal resonance at frequencies $\bar{\omega}_p$ and $\bar{\omega}_s$, while in the BIC configuration the Q-factor approaches its intrinsic value, corresponding to a much narrower estimated full width at half maximum (FWHM) of $\thicksim 74$ MHz. 
The acquired idler intensities are shown in Fig.~\ref{fig:stFWM}(e).\\ 
The \emph{equal-Q} configuration (Fig.~\ref{fig:stFWM}(e), right) exhibits a well-defined peak with a FWHM of approximately 0.7 GHz. This value is in good agreement with the expected linewidth of \mbox{$\sim 0.84\,\textrm{GHz}$}, obtained by convolving the 200 MHz pump spectrum with the product of the signal and idler cavity resonances. The latter has a FWHM of 0.64 GHz, and the convolution broadens the overall linewidth to approximately 0.84 GHz.
On the other hand, the BIC configuration (Fig.~\ref{fig:stFWM}(e), left) shows an asymmetric lineshape, where an intense and narrow-band peak with a FWHM of $0.35$ GHz sits on weaker and broader background with a FWHM of \mbox{$\thicksim1.8\,\textrm{GHz}$}. 
This pattern can be interpreted as follows: for each signal frequency $\omega_s$, energy-conservation implies that the idler photons can be generated at $\omega_i=2\bar{\omega}_p-\omega_s$, and the efficiency of this process can be enhanced by a triple-resonance condition, where all the four waves experience the maximum field enhancement. 
In the BIC condition, this occurs only at $2\bar{\omega}_p-\omega_s\thicksim\omega_{\textrm{BIC}}$, which is manifested by the spectrally narrow peak in Fig.~\ref{fig:stFWM}(e). On the other hand, when 
$|\omega_i-\omega_{\textrm{BIC}}|\gg\omega_{\textrm{BIC}}$, the field enhancement of the bound resonator mode (Fig.~\ref{fig:BIC_theo}(c)) at the idler wavelength is lost, and replaced by the lower intensity waveguide continuum of modes in Fig.~\ref{fig:BIC_theo}(b). This process creates the broad background in Fig.~\ref{fig:stFWM}(b), which tracks the form of the field enhancement of only the seed wave, still close to the expected value of $1.5$ GHz, following the same procedure. The asymmetric lineshape arises from the interference between light resonantly generated in bound mode of the main resonator with that generated into the continuum of waveguide modes. This interpretation is further supported  by the numerical simulations shown in Fig.~\ref{fig:simulations}(c) of Appendix \ref{apx:A}.

\subsection{BIC-assisted spontaneous four-wave mixing  \label{subsec:BIC-SFWM}}
\noindent
In this section, we investigate BIC-assisted spontaneous four-wave mixing for photon-pair generation, focusing on the particular case in which one of the two photons (hereby labeled as the \textit{idler}) is generated in a BIC. 
We examine three key performance metrics: the photon-pair generation rate, the second-order cross-correlation between the signal and idler modes, and the second-order self-correlation of the signal mode. 
We investigate how these quantities evolve as the idler mode transitions from the continuum (Fig.~\ref{fig:BIC_theo}(a)) to the BIC state (Fig.~\ref{fig:BIC_theo}(c)).
The pump pulse duration is fixed at 5 ns, with a repetition rate of 2 MHz and an average power of 0.5 mW on-chip. 
The use of pulsed excitation prevents the onset of thermal bistability at high pump powers while mitigating the generation of noise photons due to spontaneous Raman scattering. Background photons at the signal and idler wavelengths are further suppressed before entering the chip by means of a bandpass filter (BPF).
As shown in Fig.~\ref{fig:setup}, the generated signal and idler photons are extracted from the chip and routed to a dense wavelength division multiplexing (DWDM) filter, which suppresses the residual pump light and directs the two beams to two superconducting nanowire single-photon detectors (SNSPDs). 
Their arrival times, referenced to the electronic pump trigger, are recorded using a time-tagging unit.

\begin{figure*}[ht]
    \centering
    \includegraphics[width=\linewidth]{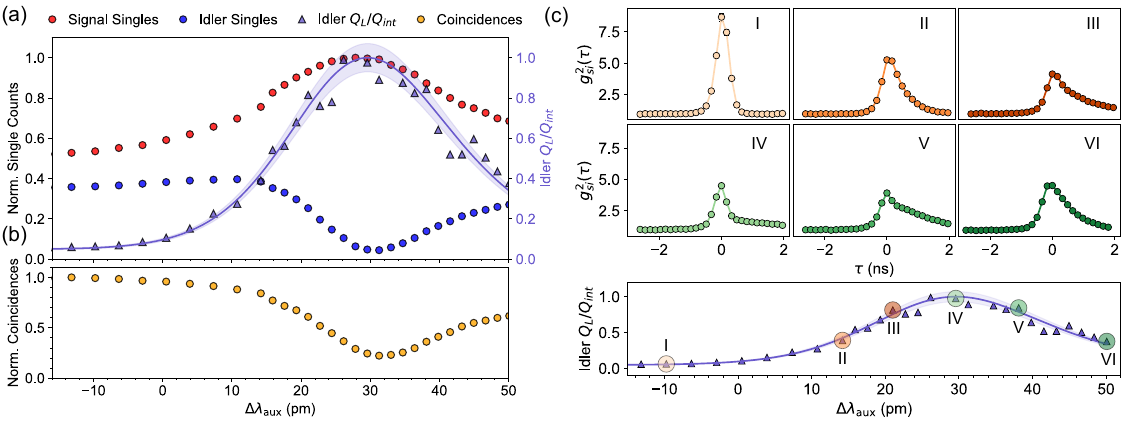}
    \caption{(a) Measured single-photon counts generated by SFWM in the signal (red) and idler (blue) modes as a function of the main-Aux. resonator detuning, $\Delta\lambda_{\mathrm{aux}}$. Signal counts are normalized to their maximum, while idler counts are normalized to the signal counts. The triangles indicate the ratio between the loaded quality factor, $Q_{\textrm{L}}$, of the idler resonance and the intrinsic quality factor of the resonator, $Q_{\textrm{int}}$. The solid line is included only as guide to the eye. (b) Normalized coincidence counts between the signal and idler modes as a function of $\Delta\lambda_{\mathrm{aux}}$. 
(c) Measured second-order signal-idler cross-correlation function, $g_{si}^{(2)}(\tau)$, as a function of the relative photon arrival time, $\tau$. The six panels, labeled I to VI, correspond to the different values of $Q_{\textrm{L}}/Q_{\textrm{int}}$ indicated in the bottom panel of (c). The lines connecting the data points are included only as guides to the eye. }
\label{fig:SinglesAndCross}
\end{figure*}

The first metric that we investigate is the photon-pair generation rate. 
Using the MZI interferometer, we tune the Q-factors of the pump and signal resonances to $\thicksim 2 \times 10^5$, corresponding to the overcoupled regime. 
We then sweep $\Delta\lambda_{\mathrm{aux}}$ from $-15$ pm to $50$ pm to selectively change the Q-factor of the idler resonance, while simultaneously recording the single-photon counts in both the signal and idler channels, their coincidence counts, and the spectrum of the idler resonance by interleaving the measurement with a wavelength sweep of an auxiliary laser. 
From the measured spectra, we extract the loaded Q-factor of the idler resonance as a function of $\Delta\lambda_{\mathrm{aux}}$. 
The results are reported in Fig.~\ref{fig:SinglesAndCross}(a), together with the single-photon counts in the signal and idler channels. 
For clarity, the coincidence counts are shown separately in Fig.~\ref{fig:SinglesAndCross}(b).\\
The BIC condition is reached when $Q_{\textrm{L}}/Q_{\textrm{int}}=1$, corresponding to $\Delta\lambda_{\mathrm{aux}}\thicksim30$ pm. 
As $\Delta\lambda_{\mathrm{aux}}$ is varied from $-15$ pm to $30$ pm, the loaded quality factor increases monotonically from $Q_\textrm{L}\thicksim2\times10^5$ to   \mbox{$Q_{\textrm{int}}=2.6\times10^6$}, passing through the critically coupled condition at $\Delta\lambda_{\mathrm{aux}}\thicksim17$ pm. 
As $\Delta\lambda_{\mathrm{aux}}$ is further increased from 30 pm to 50 pm, $Q_\textrm{L}$ decreases, and the resonator becomes critically coupled again at $\Delta\lambda_{\mathrm{aux}}\thicksim 45$ pm.
At the BIC condition, the idler single-photon counts decrease by nearly one order of magnitude ($\thicksim9$ dB) compared to the overcoupled regime at $\Delta\lambda_{\mathrm{aux}}=-10$ pm, while the coincidence counts are reduced by approximately $7$ dB. 
This behavior arises because idler photons generated in the BIC mode are decoupled from the bus waveguide and therefore cannot couple out through the output port. 
The residual single and coincidence counts originate partly from photon-pairs generated within the MZI interferometer in the asymptotic field orthogonal to the BIC mode at frequency  $\omega_{\mathrm{BIC}}$ (Fig.~\ref{fig:BIC_theo}(b)), which still exhibits a finite spatial overlap with the asymptotic field distributions of the pump and signal modes, and partly from SFWM occurring in the access bus waveguide.
To confirm that the reduction in the idler and coincidence counts is a consequence of the BIC becoming decoupled from the bus waveguide, rather than of a suppression of the SFWM process itself, we note that the signal counts roughly double  at the BIC condition relative to their value at $\Delta\lambda_{\mathrm{aux}}=-10$ pm. 
This observation demonstrates that the internal photon-pair generation rate is, in fact, enhanced under the BIC condition. 
This enhancement can be intuitively understood as a consequence of the extended lifetime of the bound idler mode, which increases the efficiency of the nonlinear interaction.\\
The second quantity that we investigate is the signal-idler second-order cross-correlation function, $g^{(2)}_{\textrm{si}}(\tau)$, where $\tau$ denotes the relative arrival time of the signal and idler photons at the detectors. 
Figure~\ref{fig:SinglesAndCross}(c) displays six representative cross-correlation curves, labeled I to VI, corresponding to the different values of $Q_\textrm{L}/Q_{\textrm{int}}$ shown in the bottom panel of the same figure. 
All the curves exhibit the characteristic double-sided, asymmetric exponential profile expected for cavity-enhanced SFWM \cite{wang2020doubly}, resulting from the unbalanced Q-factor of the signal and idler modes.
As the $Q$-factor of the idler resonance is increased (configurations from I to III), the dwelling time of the idler photons increases accordingly, resulting in progressively longer exponential tails for $\tau>0$ (falling edge), while the lifetime of the signal photons ($\tau<0$, rising edge) remains essentially unchanged. 
At the same time, both the coincidence counts and the peak value of $g^{(2)}_{\textrm{si}}(\tau)$ decrease.
At the BIC condition (configuration IV), the idler photons generated in the bound mode at $\omega_i=\omega_{\textrm{BIC}}$ are ideally decoupled from the bus waveguide. 
However, as discussed in Section~\ref{subsec:probe}, a small fraction leaks into the bus waveguide due to the finite extinction ratio of the MZI. 
In addition, photon pairs generated into the continuum of waveguide  modes at $\omega\neq\omega_{\textrm{BIC}}$ give rise to a spectrally broad background in the output waveguide (see Fig.~\ref{fig:stFWM}(e), left panel, for the classical counterpart based on stimulated emission \cite{liscidini2013stimulated}). 
These idler photons escape rapidly through the bus waveguide, whereas the corresponding signal photons continue to circulate within the resonator. 
The coexistence of photon-pair emission through the long-lived BIC mode and into the broadband continuum results in a double-exponential decay of $g^{(2)}_{\textrm{si}}(\tau)$ for $\tau>0$, where the slow and fast decay constants correspond to resonant and non-resonant idler-photon generation, respectively.
Overall, the observed trends in the single-photon counts, coincidence counts, and second-order cross-correlation shown in Fig.~\ref{fig:SinglesAndCross} are in good qualitative agreement with the numerical simulations presented in Fig.~\ref{fig:simulations}(b) of Appendix \ref{apx:A}.\\
\begin{figure}[ht]
    \centering
    \includegraphics[width=1\linewidth]{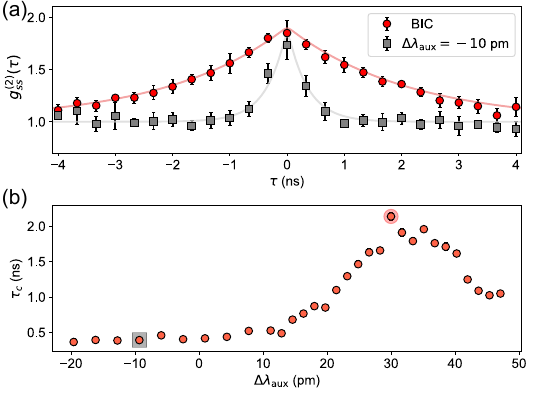}
    \caption{(a) Normalized second-order self-correlation function of the signal mode, $g^{(2)}_{\textrm{ss}}(\tau)$, as a function of the photon delay time $\tau$ for the BIC and $\Delta\lambda=-10$ pm configurations. 
    The scatter points represent the experimental data, while the solid lines show the best fits (see the main text for details). 
    (b) Signal-photon coherence time $\tau_c$ extracted from fits of $g^{(2)}_{\textrm{ss}}(\tau)$ as a function of the Aux. resonator detuning, $\Delta\lambda_{\mathrm{aux}}$. 
    The corresponding variation of the loaded $Q$-factor with $\Delta\lambda_{\mathrm{aux}}$ is shown in the bottom panel of Fig.~\ref{fig:SinglesAndCross}(c). 
    The red and gray points highlight the two measurements of $g^{(2)}_{\textrm{ss}}(\tau)$ shown in panel (a).
}
    \label{fig:selfg2}
\end{figure}
Finally, we experimentally investigate the second-order self-correlation function of the signal beam, $g^{(2)}_{\textrm{ss}}(\tau)$, by inserting a fiber beam splitter in the signal path to implement a standard Hanbury Brown and Twiss configuration. 
As in the previous measurements, we vary the loaded $Q$-factor of the idler resonance through the Aux. resonator. 
Figure~\ref{fig:selfg2}(a) shows two representative measurements of $g^{(2)}_{\textrm{ss}}(\tau)$, acquired at $\Delta\lambda_{\mathrm{aux}}=-10$ pm (overcoupled idler resonance with $Q_{\textrm{L}}=2\times10^5$) and at $\Delta\lambda_{\mathrm{aux}}=30$ pm, corresponding to the BIC condition. 
As expected, both curves are symmetric about $\tau=0$, where they reach $g^{(2)}_{\textrm{ss}}(0)=2$, reflecting the thermal statistics of the individual modes of a two-mode squeezed state \cite{christ2011probing}.
For each value of $\Delta\lambda_{\mathrm{aux}}$, we extract the coherence time $\tau_c$ by fitting the measured $g^{(2)}_{\textrm{ss}}(\tau)$ with the convolution of the exponential decay function $Ae^{-|\tau|/\tau_c}+B$, where $A$ and $B$ are free fitting parameters, and a Gaussian function with a FWHM of $99$ ps, accounting for the combined timing jitter of the two SNSPDs of around $70$ ps each. 
The extracted coherence times are reported in Fig.~\ref{fig:selfg2}(b).
Comparing the trend in Fig.~\ref{fig:selfg2}(b) with the bottom panel of Fig.~\ref{fig:SinglesAndCross}(c) reveals a strong correlation between the signal photon coherence time and the Q-factor of the partner idler mode. 
The coherence time reaches a maximum value of $\tau_{\mathrm{max}}=2.14\pm0.06$ ns at the BIC condition, in excellent agreement with the expected value of $\tau_{\mathrm{max}}=Q_{\textrm{int}}/\omega_{\mathrm{BIC}}=2.13$ ns.
Remarkably, at the BIC condition, $\tau_c$ exceeds the signal-photon lifetime, set by the loaded $Q$-factor of the cavity, by nearly one order of magnitude, reflecting the longer coherence time of its twin idler photon. 
This induced coherence admits a simple interpretation in the spectral domain. 
In a SFWM process driven by a CW pump, the triple resonance condition implies that the linewidths of the generated signal and idler photons are lower bounded by $\min(\gamma_S,\gamma_I)$, where $\gamma_{S(I)}=\frac{\bar{\omega}_{s(i)}}{Q_{L_{s(i)}}}$ denote the linewidths of the signal (idler) cavity resonances. 
In the conventional configuration of a single resonator coupled to a bus waveguide, $Q_{\textrm{L}_\textrm{s}}\thicksim Q_{\textrm{L}_\textrm{i}}$, and therefore the corresponding coherence times, $\tau_{s(i)}=\frac{Q_{\textrm{L}_{\textrm{s(i)}}}}{\omega_{s(i)}}$, are nearly identical.
In contrast, the ring-assisted interferometric coupler enables the $Q$-factors at the signal and idler wavelengths to be engineered independently. 
In the most asymmetric configuration, corresponding to the BIC condition, $\gamma_I\ll\gamma_S$ is limited only by the intrinsic loss of the cavity. 
Consequently, the signal photons are generated within a spectral bandwidth that is much narrower than the linewidth of the signal cavity resonance, thereby inheriting both the spectral bandwidth and the corresponding coherence time of the idler photons.

\subsection{BIC-assisted Optical Parametric Amplification and Oscillation \label{subsec:OPA}}
\noindent
The very high $Q$-factor of the BIC mode offers significant advantages for nonlinear parametric processes \cite{lei2023hyperparametric, wang2020doubly, ChavezBoggio2022}. 
Here, we investigate optical parametric amplification and optical parametric oscillation, demonstrating that, for a given pump power, the presence of a BIC mode leads to higher signal gain and a lower oscillation threshold compared to the \emph{equal-Q} configuration.

\begin{figure}[ht]
    \centering
    \includegraphics[width=1\linewidth]{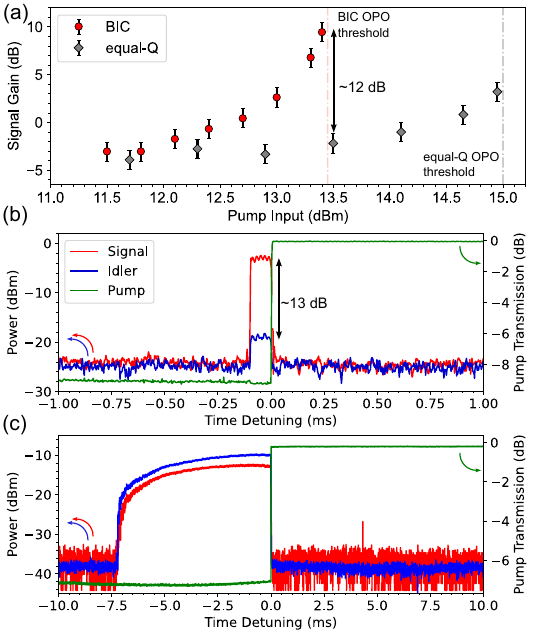}
    \caption{(a) OPA signal gain as a function of the on-chip pump power. The red and gray data points correspond to measurements acquired with the device operating in the BIC (the idler mode is a BIC) and \emph{equal-Q} configurations, respectively. The dashed vertical lines indicate the corresponding OPO thresholds. 
(b) Signal (red), idler (blue), and pump (green) intensities as a function of the laser sweep time, during which the pump wavelength is continuously increased. The abrupt change in the normalized pump transmission at $t=0$ ms results from thermal bistability. The rapid increase in the signal and idler intensities at slightly negative times marks the onset of OPO. This panel corresponds to the BIC configuration.
(c) Same as in panel (b), but for the \emph{equal-Q} configuration}
    \label{fig:OPA}
\end{figure}
These experiments were conducted under CW pumping, with the loaded $Q$-factors of the pump and signal resonances fixed at $\thicksim6\times10^5$. 
We first present the measurements related to OPA, for which the signal mode is seeded with a low input power of $-43$ dBm ($50$ nW) to prevent gain saturation. 
The pump laser wavelength is gradually tuned into resonance from the blue-detuned side, using thermal locking to stabilize the intracavity power. 
The seed laser wavelength is then swept across the signal resonance by means of the piezoelectric actuator of the external laser cavity, while the light transmitted by the resonator at the signal and idler wavelengths is recorded by two independent photodiodes connected to an oscilloscope. 
A representative measurement is shown in Fig.~\ref{fig:OPAmeas} in Appendix \ref{apx:OPA}. The signal gain, $G_{\textrm{dB}}=10\log_{10}\left(\frac{s_{\textrm{out}}}{s_{\textrm{in}}}\right)$, is extracted by comparing the off-resonance transmission, $s_{\textrm{in}}$, with the transmitted power, $s_{\textrm{out}}$, at the seed wavelength that maximizes the intensity of the stimulated idler beam (yellow dot in Fig.~\ref{fig:OPAmeas} of Appendix \ref{apx:OPA}).

Figure~\ref{fig:OPA}(a) shows $G_{\textrm{dB}}$ as a function of the on-chip pump power in the \emph{equal-Q} and BIC configurations. 

A comparison of the two curves clearly shows that the BIC configuration consistently outperforms the \emph{equal-Q} configuration, providing a higher signal gain over the entire range of pump powers. For example, at an on-chip pump power of $13.5$ dBm, the BIC-assisted OPA achieves a signal gain that is $\thicksim12$ dB higher than that of the \emph{equal-Q} configuration. We support the experimental observations with numerical simulations based on standard temporal coupled mode equations, which are adapted from \cite{zhao2023large} by including asymmetric escape efficiencies between the signal and the idler mode. The small-signal gain curves obtained from the numerical simulations, reported in Fig.~\ref{fig:OPAmeas}(c) of Appendix \ref{apx:OPA}, are in good quantitative agreement with those in the experiment.

The two vertical lines in Fig.~\ref{fig:OPA}(a) indicate the onset of OPO in the two configurations, revealing a reduction of approximately $1.6$ dB in the oscillation threshold for the BIC-assisted case.

Figures~\ref{fig:OPA}(b,c) show the intensities of the signal and idler beams as the pump wavelength is gradually tuned into resonance from the blue-detuned side at the OPO threshold (the seed laser is switched off during these measurements). In the BIC configuration, shown in Fig.~\ref{fig:OPA}(b), the signal beam is approximately $13$ dB stronger than the idler beam because it has a higher escape efficiency. In contrast, in the \emph{equal-Q} configuration shown in Fig.~\ref{fig:OPA}(c), the signal and idler beams exhibit nearly identical intensities, in agreement with their equal escape efficiencies.

\section{Discussion and conclusions}
\label{sec:discussion}
\noindent
In this work, we demonstrated a reconfigurable integrated silicon nitride photonic device in which a selected main-ring resonance approaches a Friedrich-Wintgen-type interference-induced BIC, with measured quality factors exceeding $10^6$. 
The device extends the selective linewidth control previously demonstrated with resonant interferometric couplers \cite{Paula2025,borghi2024} toward vanishing external coupling of a selected  mode.

A central result is that suppressing linear access to the idler, which remains spatially localized in the main ring and reaches an intrinsic-loss-limited lifetime, does not remove it from the nonlinear dynamics.
Indeed, we show that four-wave mixing can nevertheless generate the idler field directly inside the nonlinear main ring.
In the stimulated experiment, scattered-light imaging reveals idler light circulating in the ring, while the idler power reaching the output waveguide is reduced by approximately $22\,\mathrm{dB}$ relative to the \emph{equal-Q} configuration.
This provides an integrated $\chi^{(3)}$ realization of the broader principle that a state inaccessible through linear scattering can still be populated through a nonlinear multiphoton process \cite{calajo2019exciting}.
The resonant Aux. ring makes this cancellation wavelength selective, allowing the idler to approach the dark-state limit while the pump and signal remain radiatively coupled and can still be efficiently driven and collected.

In the spontaneous regime, the most distinctive consequence is the control exerted by the nearly dark idler mode over the marginal spectral bandwidth and temporal coherence of the extracted signal photon.
Since the joint spectral amplitude of a cavity-enhanced SFWM pair contains both cavity responses, narrowing the idler response restricts the range of signal frequencies participating in pair generation even when the idler is not detected.
Accordingly, the signal-idler cross-correlation develops strongly asymmetric exponential tails, while the measured signal coherence time increases as the idler approaches the BIC condition and reaches $2.1\,\mathrm{ns}$.
This value is consistent with the intrinsic-loss-limited lifetime of the idler mode and is almost one order of magnitude longer than the lifetime associated with the loaded signal resonance.
The long-lived idler response is therefore transferred to, and read out through, a signal photon that remains coupled to the bus waveguide, separating marginal-bandwidth control from direct extraction.
Unlike previous nonlinear microring demonstrations that primarily exploited BIC confinement to enhance conversion \cite{wang2020doubly,ye2024integrated,lei2023hyperparametric,ye2025sum}, here the nearly dark mode engineers an observable property of its radiative partner.\\
Relative to the overcoupled idler configuration at $\Delta\lambda_{\mathrm{aux}}=-10\,\mathrm{pm}$, the signal singles approximately double at the BIC condition, whereas the extracted idler singles and signal-idler coincidences decrease by approximately $9\,\mathrm{dB}$ and $7\,\mathrm{dB}$, respectively.
These opposite trends are consistent with enhanced internal SFWM accompanied by suppressed idler extraction, although Raman scattering and SFWM in the access waveguide prevent an absolute calibration of the internal pair-generation rate.\\

The OPA and OPO measurements show that the same asymmetric loss engineering remains useful in the stimulated and above-threshold regimes.
At an on-chip pump power of $13.5\,\mathrm{dBm}$, selective reduction of the idler extrinsic loss increases the measured signal gain by approximately $12\,\mathrm{dB}$ relative to the \emph{equal-Q} configuration and lowers the oscillation threshold by approximately $1.6\,\mathrm{dB}$.
Together with the spontaneous measurements, these results show that mode-selective dissipation engineering provides a common strategy across the quantum, stimulated, and oscillating regimes.\\
The current implementation is limited by residual point-coupler imbalance, intrinsic propagation loss, and background photons generated in the access waveguide.\\

Looking ahead, selecting which resonances become nearly dark while leaving the other modes accessible could enable electrically tunable photon bandwidths, higher-gain parametric amplifiers, and lower-threshold oscillators.
Overall, these results establish wavelength-selective dissipation engineering as a route to controlling nonlinear interaction, spectral shaping, and photon extraction as distinct degrees of freedom within the same integrated device.

\section{ACKNOWLEDGMENTS}
\noindent
E.C.L acknowledges the funding of Fundação de Amparo à Pesquisa do Estado de São Paulo (FAPESP) (Grant No. 2025/01292-2).
D.B. acknowledges European Union funding from the STARLight project (project ID: 101194170). M.C., M.L., M.G. and M.B. acknowledge the PNRR MUR project PE0000023-NQSTI.
M.C. acknowledges funding from the European Union under the MSCA Postdoctoral Fellowship grant 101211100 (project GLINT).
\appendix
\section{Numerical simulations of photon-pair generation in the photonic device \label{apx:A}}
In this section, we present numerical simulations of SFWM in the photonic device to support the experimental results and provide further insight into the role of BICs in photon-pair generation.

\begin{figure}[ht]
\centering
\includegraphics[width=1\linewidth]{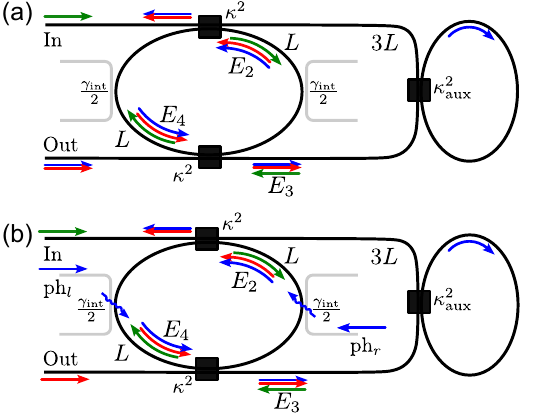}
\caption{(a) Sketch of the asymptotic fields associated with the pump entering through the input port and the signal and idler photons entering through the output port. 
The light-gray waveguides represent the \emph{phantom} channels associated with the intrinsic loss of the cavity. 
The fields $E_4$ and $E_2$ denote the complex amplitudes propagating in the left and right sections of the main resonator, respectively. 
The definition of the symbols is provided in the text. 
(b) Asymptotic fields associated with the  idler photons entering the left and right \emph{phantom} channels, while signal photons entering from the output port, configuration used to calculate the signal broken pairs. 
The same is done with signal photons entering the left and right \emph{phantom} channels, and the idler entering the output port.}
\label{fig:asyfields}
\end{figure}

We employ the formalism of asymptotic fields in a lossy resonator to account for the complex field distribution within the structure and the presence of multiple ports through which photons can enter and leave the nonlinear interaction region \cite{Liscidini2012,banic2022two}. Note that this approach reduces to the asymptotic-field formalism introduced in Section \ref{sec:theory} in the limit of vanishing loss. 
For clarity, the schematic of the device shown in Fig.~\ref{fig:res}(a) is reproduced in Fig.~\ref{fig:asyfields}, together with the main parameters of the model: the resonator-bus-waveguide and auxiliary-bus-waveguide coupling coefficients $(\kappa^2,\kappa_{\mathrm{aux}}^2)$, the half-perimeter of the main resonator $L$, and the intrinsic decay rate $\gamma_i$, all listed in Table~\ref{tab:simulation_parameters}.
The figure also indicates the propagation directions of the pump (green), signal (blue), and idler (red) asymptotic fields along the structure. 
Intrinsic losses are modeled by two \emph{phantom} waveguides coupled to the main resonator, each with a coupling rate of $\gamma_{\mathrm{int}}/2$, such that the total intrinsic loss is equal to $\gamma_{\mathrm{int}}$ \cite{banic2022two}.
\begin{table}[!ht]
\centering
\caption{Description and values of the parameters used for the numerical simulations of the asymptotic fields and in the coupled wave Eqs.(\ref{eq:coupled_modes}). 
The values of the entries denoted with a $(^{*})$ symbol have been extracted from the experiment, while the other entries have been calculated through finite element method.}
\begin{tabular}{ccc}
\hline
Parameter 1 & Description 2 & Value \\
\hline
$\kappa^{2*}$ & Main-bus coupling coefficient & $0.03$ \\
$\kappa_{\mathrm{aux}}^{2*}$ & Aux-bus coupling coefficient & $0.12$ \\
$\gamma_{\mathrm{int}}^*$ & Intrinsic loss rate & $2\pi\times74$ MHz \\
$L$ & Resonator half-perimeter  & $400 \; \mu$m \\
$\textrm{FSR}^*$ & Free Spectral Range  & $195$ GHz \\
$\gamma_{p0}^*$ & Extrinsic pump loss rate  & $2\pi\times1$ GHz \\
$\gamma_{s0}^*$ & Extrinsic signal loss rate & $2\pi\times1$ GHz \\
$\gamma_{i0}^*$ & Extrinsic idler loss rate & $2\pi\times1$ GHz \\
$\gamma_{\mathrm{nl}}$ & Nonlinear coefficient  & $138$ MHz$/$W  \\
$D_{\textrm{int}}^*$ & Integrated dispersion  & $2\pi\times250$ MHz \\
\hline
\end{tabular}
\label{tab:simulation_parameters}
\end{table}

Following Ref.~\cite{Liscidini2012}, the calculation of the SFWM generation rate and of the spectral correlations between the signal and idler photons requires determining the asymptotic field distributions of the modes participating in the nonlinear interaction. 
In particular, the field amplitudes of interest are those labeled $E_2$ and $E_4$ in Fig.~\ref{fig:asyfields}. 
These correspond to the fields circulating inside the main resonator, where photon-pair generation is assumed to occur owing to the strong resonant field enhancement.

The pump is injected through the input port, and the corresponding intracavity amplitudes $E_{2(4)p}$ are obtained using the standard scattering-matrix formalism \cite{Paula2025}. 
Signal and idler photons can leave the structure through the output port (where they can be detected) and through the two \emph{phantom} channels, where they are irreversibly lost. 
The corresponding field distributions are obtained by injecting a unit-amplitude field into the output port (Fig.~\ref{fig:asyfields}(a)) or into the left (\emph{l}) and right (\emph{r}) \emph{phantom} channels (Fig.~\ref{fig:asyfields}(b)), and solving for the corresponding intracavity fields $E^{\textrm{out}}_{2s(i)}$, $E^{\textrm{out}}_{4s(i)}$, $E^{\textrm{ph}_l}_{2s(i)}$, and $E^{\textrm{ph}_r}_{4s(i)}$, respectively.

We now define the complex function $\tilde{\phi}_{x_sy_i}(\omega_s,\omega_i)$ as
\begin{equation}
\tilde{\phi}_{x_sy_i} =  \int f(\omega_s,\omega_i,\mathbf{r})E^{x_s}_{s}(\omega_s,\mathbf{r})E^{y_i}_{i}(\omega_i,\mathbf{r})d\mathbf{r},
\label{eq:JSA}
\end{equation}

where the spatial integration is performed over the entire volume of the resonator and $x_s,y_i$ can indicate either the output waveguide ($x_s,y_i=\textrm{out}$) or the left-right \emph{phantom} channels (($x_s,y_i=(\textrm{ph}_l,\textrm{ph}_r)$)). 
The function $f(\omega_s,\omega_i,\mathbf{r})$ is defined as
\begin{align}
&f(\omega_s,\omega_i,\mathbf{r}) =\\ &\int \alpha_p(\omega)\alpha_p(\omega_s+\omega_i-\omega)E_p(\omega,\mathbf{r})E_p(\omega_s+\omega_i-\omega,\mathbf{r})d\omega. \nonumber
\end{align}

The quantity $\alpha_p(\omega)$ is the normalized, square-integrable pump spectral envelope centered at the pump resonance frequency $\bar{\omega}_p$, which we model as the Fourier transform of a top-hat pulse of duration $T$ in the time domain,
\begin{equation}
\alpha_p(\omega) = T\textrm{sinc}\left(\frac{T(\omega-\bar{\omega}_{p})}{2}\right).
\end{equation}

The functions $E^{x_s(y_i)}_{s(i)}(\omega_{s(i)},\mathbf{r})$ are defined such that $E^{x_s(y_i)}_{s(i)}(\omega_{s(i)},\mathbf{r})=E^{x_s(y_i)}_{2s(i)}(\omega_{s(i)},\mathbf{r})$ when the position vector $\mathbf{r}$ points the half-perimeter of length $L$ on the right side of the point-coupling region, and $E^{x_s(y_i)}_{s(i)}(\omega_{s(i)},\mathbf{r})=E^{x_s(y_i)}_{4s(i)}(\omega_{s(i)},\mathbf{r})$ when $\mathbf{r}$ points the half-perimeter of length $L$ on the left side of the point-coupling region. 
The pump field $E_p(\omega,\mathbf{r})$ is defined analogously. 
Note that the dependence of the device response on $\Delta\lambda_{\mathrm{aux}}$ is entirely contained in the expressions for the internal fields $E^{x_s(y_i)}_{s(i)}$ and $E_p$.

The function $\tilde{\phi}_{x_sy_i}$ allows the coincidence rate of signal and idler photons exiting ports $x_s$ and $y_i$, respectively, to be calculated as
\begin{equation}
C_{x_sy_i}=\bar{R}\int |\tilde{\phi}_{x_sy_i}(\omega_s,\omega_i)|^2d\omega_s,d\omega_i,
\label{eq:coincidence_rate}
\end{equation}

where $\bar{R}$ is a constant that depends on the material nonlinearity and the effective waveguide mode area \cite{lukens2026beyond}. 
Since our interest is limited to normalized quantities that highlight the dependence of $C_{x_sy_i}$ on $\Delta\lambda_{\mathrm{aux}}$, we do not explicitly evaluate $\bar{R}$.

Similarly, the signal and idler photon rates exiting port $x_s$ and $y_i$ are obtained as $R_{x_s}=\sum_{y_i} C_{x_sy_i}$  and $R_{y_i}=\sum_{x_s} C_{x_sy_i}$. 
In this work, we are interested in the single-photon and coincidence rates corresponding to signal and idler photons exiting through the output waveguide. 
Accordingly, we set $x_s=y_i=\mathrm{out}$ and let the sum to run over $\{\mathrm{out},\mathrm{ph}_l,\mathrm{ph}_r\}$, thereby marginalizing over all possible exit ports of the undetected photon.

To compare the numerical simulations with the experimental observations, we first calculate the transmission spectra of the device around the idler resonance as a function of $\Delta\lambda_{\mathrm{aux}}$. 
This preliminary step allows us to identify the value of $\Delta\lambda_{\mathrm{aux}}$ at which the main resonator becomes effectively decoupled from the access waveguide, thereby supporting a BIC.

\begin{figure}[ht]
    \centering
    \includegraphics[width=1\linewidth]{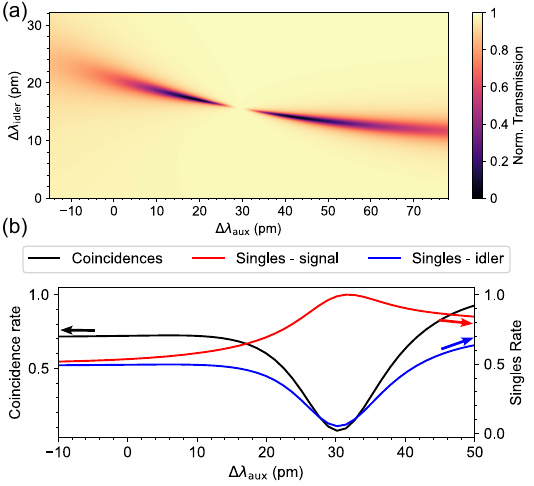}
    \caption{(a) Simulated transmission spectra (stacked vertically) of the device around an idler resonance as a function of $\Delta\lambda_{\mathrm{aux}}$. The detuning $\Delta\lambda_i=\lambda-\lambda_i$ is defined with respect to the idler resonance wavelength $\lambda_i$, determined for $\Delta\lambda_{\mathrm{aux}}$ much larger than the FWHM of the Aux. resonator (i.e., the unperturbed resonance wavelength). (b) Simulated singles count rates and signal-idler coincidence rate as a function of $\Delta\lambda_{\mathrm{aux}}$. The singles count rates of both the signal and idler channels are normalized to the maximum signal count rate, while the coincidence rate is normalized to one.}
    \label{fig:single-sim}
\end{figure}

Figure~\ref{fig:single-sim}(a) shows the stacked transmission spectra of the idler resonance as a function of $\Delta\lambda_{\mathrm{aux}}$. 
At \mbox{$\Delta\lambda_{\mathrm{aux}}\thicksim30\,\textrm{pm}$}, the extinction ratio vanishes, indicating that the resonator is effectively decoupled from the access bus waveguide at the idler wavelength. 
This value is in close agreement with the experimental result shown in Fig.~\ref{fig:SinglesAndCross}(a). 
The transmission reaches zero at \mbox{$\Delta\lambda_{\mathrm{aux}}\thicksim15\,\textrm{pm}$} and $\Delta\lambda_{\mathrm{aux}}\thicksim45$ pm, corresponding to the critically coupled condition. 
Away from these points, the resonance broadens and the transmission increases, marking the transition to the overcoupled regime. 
The transmission spectrum is not symmetric about the BIC condition because the Aux. resonator resonance has a different detuning with respect to the corresponding resonance of the main resonator.
\begin{figure*}[ht]
    \centering
    \includegraphics[width=1\textwidth]{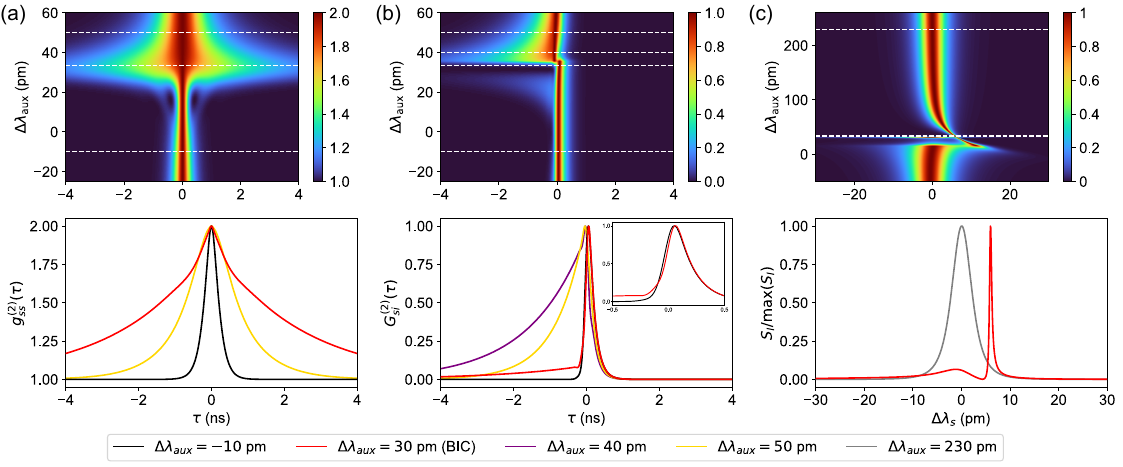}
    \caption{(a) Simulated second-order self-correlation function $g_{ss}^{(2)}(\tau)$ of the signal beam as a function of the time delay $\tau$ for different Aux. resonator detunings $\Delta\lambda_{\mathrm{aux}}$. (b) Normalized cross-correlation function $G_{si}^{(2)}(\tau)$ between the signal and idler beams. The inset shows a magnified view of the region $-0.5 \leq \tau \leq 0.5$, highlighting the double-exponential left tail of the $G_{si}^{(2)}(\tau)$ function close to the BIC condition at $\Delta\lambda_{\mathrm{aux}}=30$ pm. (c) Normalized StFWM intensity (Eq.(\ref{eq:StFWM_spectra})) of the idler beam as a function of the wavelength detuning $\Delta\lambda_s$ of the seed laser with respect to the main ring resonance. The $\Delta\lambda_{\mathrm{aux}}=230$ pm refers to the \emph{equal-Q} configuration in Fig.\ref{fig:stFWM}(e). In both panels (b) and (c), the curves are normalized to one for each value of $\Delta\lambda_{\mathrm{aux}}$ for clarity. All the curves are convolved to the gaussian-shaped detector response, of $99$ ps FWHM. The curves in the lower panels correspond to the horizontal cross sections, marked by dashed white lines, of the respective upper panels.}
    \label{fig:simulations}
\end{figure*}
Figure ~\ref{fig:single-sim}(b) shows the simulated normalized single rates $R_{\textrm{out}_\textrm{s}}$ and $R_{\textrm{out}_\textrm{i}}$, along with the coincidence rate $C_{\textrm{out}_\textrm{s},\textrm{out}_\textrm{i}}$. The trends predicted by the simulation is in qualitative agreement with the experimental data in Fig.~\ref{fig:SinglesAndCross}(a). Notably, the model accurately predicts the reduction in idler rates and the enhancement of the signal rate in proximity to the BIC condition. One of the main discrepancies between the experimental results in Fig.~\ref{fig:SinglesAndCross}(a) and the simulations in Fig.~\ref{fig:single-sim}(b) is that they predict a much severe reduction of both the coincidences and the idler counts at the BIC condition. We attribute the discrepancy to noise photons generated by spontaneous Raman scattering and SFWM in the bus waveguide, which are not modeled in the simulations but clearly impacting the experimental measurements (see e.g. the noise limited coincidence to accidental ratio $\textrm{CAR}=g^{(2)}_{\textrm{si}}-1$ shown in Fig.~\ref{fig:SinglesAndCross}(c)). 
We then turn our focus to the calculation of the self and cross-correlations $g_{\textrm{ss}}^{(2)}(\tau)$ and $g_{\textrm{si}}^{(2)}(\tau)$.
The self correlation $g^{(2)}_{\textrm{ss}}(\tau)$ is calculated by exploiting the relation $g_{\textrm{ss}}^{(2)}(t_s,t_i) = 1 + |g_{\textrm{ss}}^{(1)}(t_s,t_i)|^2$, valid for thermal states, where
\begin{equation}
    g_{\textrm{ss}}^{(1)}(t_1,t_2)=\frac{\langle a_s(t_1)^{\dagger}a_s(t_2) \rangle}{\sqrt{\langle a_s(t_1)^{\dagger}a_s(t_1)\rangle \langle a_s(t_2)^{\dagger}a_s(t_2)\rangle}}, \label{eq:g_1}
\end{equation}
and $a_s(t)$($a_s(t)^{\dagger}$) are the annihilation(creation) operators of a signal photon at time $t$.
It is straightforward to show that $\langle a_s^{\dagger}(t_1)a_s(t_2)\rangle=\mathcal{F}(\rho_s(\omega_s,-\omega_s'))$, where $\mathcal{F}$ denotes the two dimensional Fourier transform and $\rho_s$ contains the elements of the one-photon reduced density matrix of the signal, which is expressed in the frequency basis as
\begin{equation}
    \rho_s(\omega_s,\omega_s')=\beta\sum_{y_i}\int \phi_{\textrm{out}_s,y_i}(\omega_s,\omega_i)\phi_{\textrm{out}_s,y_i}^{*}(\omega_s',\omega_i)d\omega_i,  
    \label{eq:red_density_matrix}
\end{equation}
where $\beta$ is the photon-pair generation probability and  $\phi_{x_sy_i}(\omega_s,\omega_i)\propto\tilde{\phi}_{x_sy_i}(\omega_s,\omega_i)$ is the biphoton wavefunction, whose modulus square describes the density probability of generating a signal photon at frequency $\omega_s$ in port $x_s$ and an idler photon at frequency $\omega_i$ in port $y_i$. Similarly, the cross-correlation function $g^{(2)}_{\textrm{si}}(t_1,t_2)$ is defined as
\begin{equation}
    g_{\textrm{si}}^{(2)}(t_1,t_2)=\frac{\langle a_s(t_1)^{\dagger}a_i(t_2)^{\dagger}a_s(t_1)a_i(t_2) \rangle}{\langle a_s(t_1)^{\dagger}a_s(t_1)\rangle \langle a_i(t_2)^{\dagger}a_i(t_2)\rangle}, \label{eq:g_2si}    
\end{equation}
and the numerator can be expressed as
\begin{equation}
\langle a_s(t_1)^{\dagger}a_i(t_2)^{\dagger}a_s(t_1)a_i(t_2) \rangle=|\mathcal{F}(\phi_{\textrm{out}_s,\textrm{out}_i}(\omega_s,\omega_i))|^2 . \label{eq:cross_g2}
\end{equation}
Note that, contrarily to the $g^{(1)}_{\textrm{ss}}$, the proper normalization of the cross-correlation require the explicit calculation of $\beta$. Since our main goal is to qualitatively reproduce the trends shown in Fig.~\ref{fig:SinglesAndCross}(c), in the following we focus our attention on the normalized quantity $G^{(2)}_{\textrm{si}}(\tau)=g^{(2)}_{\textrm{si}}(\tau)/\textrm{max}(g_{\textrm{si}}^{(2)})$.   
For pump pulse durations $T$ much longer than the dwelling time of photons in the cavity, the biphoton wavefunction becomes strongly anti-correlated. In the CW limit where $T\rightarrow\infty$, $\phi_{\textrm{out}_s,y_i}(\omega_s,\omega_i)=\Phi_{\textrm{out}_s,y_i}(\omega_s)\delta(\omega_i-(2\omega_p-\omega_s))$, Eq.(\ref{eq:red_density_matrix}) simplifies to $\rho_s(\omega_1,\omega_2)=\sum_{y_i}|\Phi_{\textrm{out}_s,y_i}(\omega_1)|^2\delta(\omega_1-\omega_2)$  and $\langle a_s^{\dagger}(t_1)a_s(t_2)\rangle=\langle a_s^{\dagger}(0)a_s(\tau)\rangle=\sum_{y_i}\int|\Phi_{\textrm{out}_s,y_i}(\omega)|^2e^{i\omega\tau}d\omega$. In this limit, all the  correlation functions depend only on the relative time $\tau=t_2-t_1$, as expected for a stationary process. In practice, for finite pulse durations which are much longer than the dwelling time of photons in the cavity, we discard transient behaviors at the beginning and the end of the pulse, truncating the correlation functions in a region $t_1,t_2\in(-T_c,T_c)^2$ where they have reached their stationary values,  and approximate $g^{(2)}_{\textrm{ss/si}}(\tau)$ with $g^{(2)}_{\textrm{ss/si}}(\tau)\thicksim\frac{1}{2T_c}\int_{-T_c}^{T_C} g_{\textrm{ss/si}}^{(2)}(t,t+\tau)dt$. The truncation interval is determined by increasing $T_c$ until the value of the correlation function departs from its stationary value by more than $10\%$.\\
The simulated $g^{(2)}_{\textrm{ss}}$ and $G^{(2)}_{\textrm{si}}$ as a function of $\Delta \lambda_{\mathrm{aux}}$ are displayed in Fig.~\ref{fig:simulations}(a) and Fig.~\ref{fig:simulations}(b), respectively. In Fig.~\ref{fig:simulations}(a), we highlight three cross sections, indicated by white dashed lines and plot in the lower panels, corresponding to different values of $\Delta\lambda_{\mathrm{aux}}$. By relating $\Delta\lambda_{\mathrm{aux}}$ to the transmission spectra in Fig.~\ref{fig:single-sim}(a), we see that the signal coherence time is strongly correlated with the idler spectral linewidth, in agreement with the experimental results shown in Fig.~\ref{fig:selfg2}. In particular, at the BIC condition, the coherence time is maximized. 
The simulated $G^{(2)}_{\textrm{si}}(\tau)$ as a function of $\Delta\lambda_{\mathrm{aux}}$ is shown in Fig.~\ref{fig:simulations}(b). Four cross sections, corresponding to different values of $\Delta\lambda_{\mathrm{aux}}$, are shown in the lower panel.
In close agreement with the experimental results shown in Fig.~\ref{fig:SinglesAndCross}(c), the simulations reproduce the asymmetric exponential tails corresponding to the different escape efficiencies of the signal-idler photons (see e.g. the cross sections at $\Delta\lambda =40$ pm and $\Delta\lambda =50$ pm), as well as the double exponential tail ($\Delta\lambda =30$ pm) at the BIC condition. Remarkably, the decay time of the idler photon increases from $\tau_i=300$ ps at $\Delta\lambda_{\mathrm{aux}}<0$ to $\tau_i=2.1$ ns as the system approaches the BIC regime. At the BIC point, the double exponential tail emerges, where the  short and long coherence times are $10$ ps and $2.1$ ns. This behavior is consistent with the experimental trends shown in Fig.~\ref{fig:SinglesAndCross}(c) and is discussed in detail in the main text. The shorter tail has a decay costant of the order of $\textrm{FSR}^{-1}=\frac{L}{v_g}$ (where $v_g\sim150\,\mu\textrm{m}/\textrm{ps}$ is the group velocity) and corresponds to idler photons generated in the half-perimeter of length $L$ on the right side of the point-coupling region.
Finally, we simulate the intensity of the idler beam generated by StFWM as a function of the detuning $\Delta\lambda_s$ between the seed wavelength and the signal resonance. The predicted spectra has to be compared to the corresponding experimental results shown in Fig.~\ref{fig:stFWM}(e) of the main text. The normalized StFWM idler intensity $S_{I}(\omega_s)/\textrm{max}(S_I)$, shown in Fig.~\ref{fig:simulations}(c), has been calculated by using the following expression
\begin{equation}
    S_I(\omega_s) = \int|\tilde{\phi}_{\textrm{out}_s,\textrm{out}_s}(\omega_s,\omega_i)|^2d\omega_i. \label{eq:StFWM_spectra}
\end{equation}
The values of the detuning $\Delta\lambda_{\mathrm{aux}}$ corresponding to the two device configurations shown in Fig.~\ref{fig:stFWM}(e) are \mbox{$\Delta\lambda_{\mathrm{aux}}=230$} pm (\emph{equal-Q}) and \mbox{$\Delta\lambda_{\mathrm{aux}}=30$} pm (BIC). The related spectra are shown in the bottom panel of Fig.~\ref{fig:simulations}(c). As one can see, the simulation reproduces the most relevant qualitative features observed in the experiment. First, the peak intensity of the StFWM signal red shifts by transitioning from the \emph{equal-Q} to the BIC configuration. Second, at the BIC condition, the resonant peak attributed to the bound mode is sitting on a broad signal background corresponding to the continuum of waveguide modes, in agreement with the experimental data. 

\section{OPA-gain Measurement and Simulation \label{apx:OPA}}

\begin{figure}[ht]
    \centering
    \includegraphics[width=1\linewidth]{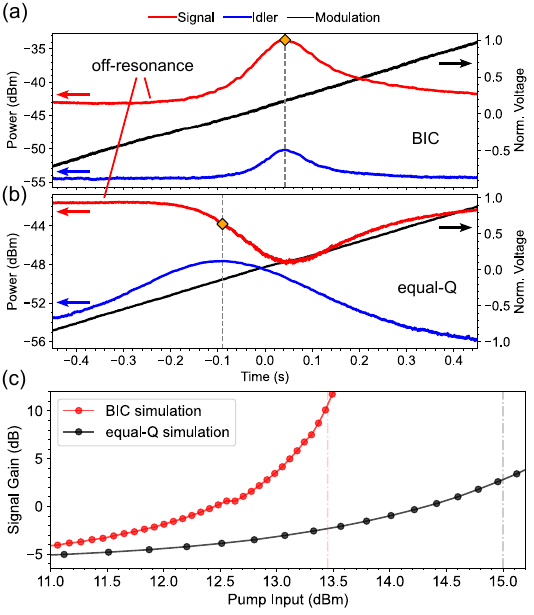}
    \caption{OPA gain characterization and simulation. (a) OPA measurement for BIC configuration at pump power around $13.5$ dBm. (b) OPA measurement for \emph{equal-Q} configuration with same pump power. Simulation of OPA-induced signal gain in both BIC and \emph{equal-Q} configurations, corresponding to the experimental measurement seen in Fig.~\ref{fig:OPA}(a). The vertical lines are the OPO threshold measured in the experimental plot, and serve as a guide to the eye.}
    \label{fig:OPAmeas}
\end{figure}

In our OPA-gain measurement procedure, the seed laser wavelength is continuously swept using its internal piezo-actuator, which is driven by a saw-tooth voltage waveform with a period of $0.5$ Hz. During this process, the pump laser is set to a constant power, and its wavelength is passively locked close to one of the resonances of the main resonator by exploiting thermal locking. Two representative examples of the time traces acquired during the wavelength sweep, measured at the same pump power of $13.5$ dBm and corresponding to the intensities of the signal and idler beams, are shown in Fig.~\ref{fig:OPAmeas}(a,b).

In the configuration shown in Fig.~\ref{fig:OPAmeas}(b), the quality factor of all three resonances involved in the OPA process is set to $6\times10^5$. Under this condition, there is no net OPA gain, and the signal transmission exhibits the characteristic resonance dip. Conversely, in the configuration shown in Fig.~\ref{fig:OPAmeas}(a), the idler mode corresponds to a BIC, resulting in a net OPA gain of $\thicksim10$ dB (see Fig.~\ref{fig:OPA}(a)). Consequently, the transmitted signal intensity is amplified with respect to the input signal, reaching its maximum value when the signal seed is resonant with the main ring. As already mentioned in the main text, the signal gain is evaluated at the seed wavelength corresponding to the maximum idler intensity. The same criterion is adopted to evaluate the signal gain from the simulations.

The simulation of the OPA gain is based on the well-established formalism of temporal coupled-mode equations describing below-threshold, degenerately pumped four-wave mixing in optical microcavities. We use as a reference the equations reported in the Supplementary Material of Ref.~\cite{zhao2023large}, which are reproduced here for clarity:

\begin{equation}
\label{eq:coupled_modes}
\begin{aligned}
\frac{dA}{dt} &= -\frac{\gamma_{A}}{2}A - i\Delta_A A
+ i\gamma_{\mathrm{nl}}(|A|^2 + 2|B|^2 + 2|C|^2)A \\
&\quad + 2i\gamma_{\mathrm{nl}} A^* BC + \sqrt{\gamma_p\mathrm{FSR}}\,A_{\textrm{in}}, \\
\frac{dB}{dt} &= -\frac{\gamma_B}{2}B - i\Delta_B B
+ i\gamma_{\mathrm{nl}}(2|A|^2 + |B|^2 + 2|C|^2)B \\
&\quad + i\gamma_{\mathrm{nl}} A^2 C^* + \sqrt{\gamma_s\mathrm{FSR}}\,B_{\mathrm{in}}, \\
\frac{dC}{dt} &= -\frac{\gamma_C}{2}C - i\Delta_C C
+ i\gamma_{\mathrm{nl}}(2|A|^2 + 2|B|^2 + |C|^2)C \\
&\quad + i\gamma_{\mathrm{nl}} A^2 B^*.
\end{aligned}
\end{equation}

In this set of equations, $A$, $B$ and $C$ corresponds to the slowly-varying temporal envelope of the pump, signal and idler modes respectively, normalized such that their modulus square have the units of power. The detuning $\Delta_C$ of the idler wave compared to its cold (no self and cross-phase modulation induced shifts) resonance frequency is constrained to satisfy the energy conservation relation $2\Delta_A - \Delta_B - \Delta_C + D_{\textrm{int}}=0$, where $D_{\textrm{int}}$ is the integrated dispersion. The total coupling rate for each mode $\gamma_{J=A,B,C}$ is described as $\gamma_J=\gamma_j+\gamma_{\mathrm{int}}$, where $\gamma_{j=p,s,i}$ is the extrinsic coupling rate. The description of the other quantities and their values are reported in Table \ref{tab:simulation_parameters}.

A key variation that we had to include in our model is the dependence of the extrinsic coupling rate with the intracavity pump power $|A|^2$, that we model to the first order as $\gamma_j = (\gamma_{j0} + \xi|A|^2)$, where $\xi=-1\times10^8$ W$^{-1}$s$^{-1}$ is a coefficient that has been determined from the experiment and $\gamma_{j0}$ is the cold cavity extrinsic coupling rate.

This dependence is introduced to take into account the fact that the internal pump power modifies the temperature of the material, which in turn changes the effective index due to the thermo-optic effect. 
Half of the perimeter of the main ring is embedded into one of the two arms of the MZI interferometer, converting temperature variations in changes of their relative phase, and ultimately to the effective coupling coefficient.

The steady state solution of the set of Eqs.(\ref{eq:coupled_modes}) is numerically solved by fixing the input seed power $|B_{in}|^2$ to $-43$ dBm, i.e., the value used in the experiment, and varying $|A_{\textrm{in}}|^2$, the pump input power. The simulated signal gain is shown in Fig.~\ref{fig:OPAmeas}(c) for both the \emph{equal-Q} and BIC configurations, indicating a close agreement with the experimental measurements shown in Fig.~\ref{fig:OPA}(a).

\bibliography{references}

\end{document}